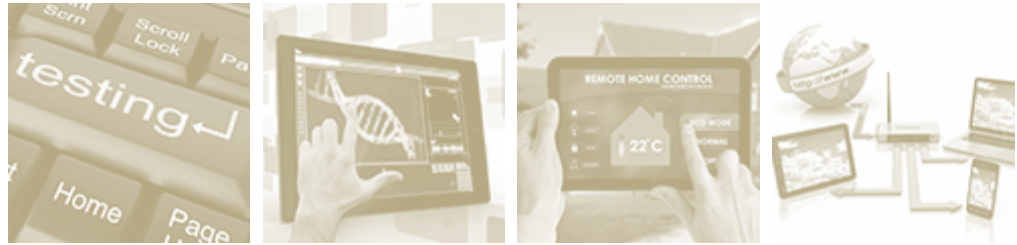

# GREAT Process Modeller user manual

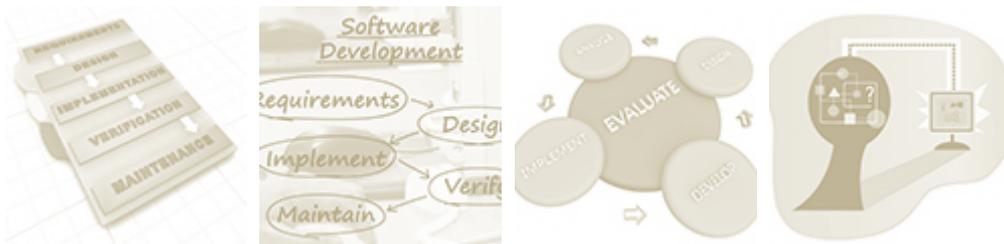

| Reference number | PROS-TR-2015-01 |
| --- | --- |
| Title | GREAT Process Modeller user manual |
| Author (s) | Urko Rueda, Sergio España, Marcela Ruiz |
| Corresponding author (s) | {urueda, sergio.espana, lruiz} @pros.upv.es |
| Document version number | 1.0    **Final version** No    **Pages** 8 |
| Release date | January 2015 |
| Key words | User manual, Requirements engineering, Communication Analysis, model-driven development, Message Structures, business process modelling, UML Class Diagrams, OO-Method |

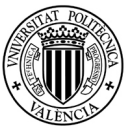

# TABLE OF CONTENTS

# 1. Scope of this document

This report contains instructions to install, uninstall and use GREAT Process Modeller[1], a tool that supports Communication Analysis, a communication-oriented business process modelling method. GREAT allows creating communicative event diagrams (i.e. business process models), specifying message structures (which describe the messages associated to each communicative event), and automatically generating a class diagram (representing the data model of an information system that would support such organisational communication).

This report briefly describes the methodological background of the tool. This handbook explains the modelling techniques in detail:

> España, S., A. González, Ó. Pastor and M. Ruiz (2012). Communication Analysis modelling techniques. Technical report ProS-TR-2012-02, PROS Research Centre, Universitat Politècnica de València, Spain, http://arxiv.org/abs/1205.0987

If, when writing a scientific publication, you intend to cite the Communication Analysis method as a whole, please use the following reference:

> España, S., A. González and Ó. Pastor (2009). Communication Analysis: a requirements engineering method for information systems. 21st International Conference on Advanced Information Systems (CAiSE'09). Amsterdam, The Netherlands, Springer LNCS 5565: 530-545.

If you intend to cite the Message Structures modelling technique in particular, please use the following reference:

> González, A., M. Ruiz, S. España and Ó. Pastor (2011). Message Structures: a modelling technique for information systems analysis and design. 14th Workshop on Requirements Engineering (WER 2011). M. Lencastre and H. Estrada. Rio de Janeiro, Brazil, extended version in English and Spanish available at http://arxiv.org/abs/1101.5341: 407-418.

If you intend to cite the model-to-model transformations, please use the following references:

> González, A., S. España, M. Ruiz and Ó. Pastor (2011). Systematic derivation of class diagrams from communication-oriented business process models. 12th Working Conference on Business Process Modeling, Development, and Support (BPMDS'11). T. A. Halpin, S. Nurcan, J. Krogstieet al. London, UK, Springer LNBIP. 81: 246-260.

> España, S., M. Ruiz, Ó. Pastor and A. González (2011). Systematic derivation of state machines from communication-oriented business process models. IEEE Fifth International Conference on Research Challenges in Information Science (RCIS 2011). Guadeloupe, France, IEEE: 1-12.

---

[1] GREAT stands for Global Reengineering Environment with Automated Transformations

## 2. Tool goal

It is a platform for business process modelling and information system' requirements specification focused to support software production and maintenance of excellence. Software crisis manifests in hard to manage projects, wasting time and money, and low-quality software, which does not satisfy customer requirements and is expensive to maintain.

The tool is **global**, covering the whole software development process (from system analysis and requirements specifications to code production and maintenance). It supports **reengineering** projects helping them in the adaptation of existing software assets to new environment needs. And, it supports agile development through **automated transformations** inside a model-driven methodology for software development.

Business architects, Software analysts, and all software development/maintenance concerned parties interested in the production and maintenance of quality software are the expected users of the tool.

## 3. Tool methodology

The platform enables the agile analyses of business processes and let analysts and interviews' participants focus in the business activities, which brings new information to business, abstracting details that do not contribute to product development. It also shortens the development, or reengineering, time applying model transformation technologies that automate part of the system design.

This is achieved by the Communication Analysis methodology, which is comprised of several modelling techniques, namely *Communicative Event Diagrams*, *Message Structures* and *Event Specification Templates*, and a transformation engine. Transformations help in the process of changing the specification perspective of information system' requirements for the intended audience. The perspective is changed from:

a) the requirements specification perspective, for users that better perform focusing on the business activities and interactions

to:

b) the system specification perspective (UML has been selected for the purpose), for users related to its development.

Thus, the methodology contributes to a better approach of software products specifications in which the most critical users (analysts, business architects, business modellers, etc.) are able to specify the systems requirements at an appropriate perspective, and obtain the related UML specifications automatically.

### 3.1. Communicative Event Diagram modelling technique

The *Communicative Event Diagram* is a business process modelling technique that adopts a communicational perspective by focusing on communicative interactions when describing the organizational work practice, instead of focusing on physical activities[2]; at this abstraction level, we refer to business activities as communicative events.

## 3.2. Message Structures modelling technique

*Message Structures* is a technique based on structured text that allows specifying the messages associated to communicative events.

## 3.3. Event Specification Templates technique

*Event Specification Templates* are a means to organise the requirements concerning a communicative event.

## 3.4. Transformation engine

Once the system requirements have been defined throughout the previous modelling techniques, the transformation engine switches the specification perspective to UML diagrams. The engine is sustained on several metamodels (Cametamodel[3], ecore, uml, tracer, todo) that provide the ability to automatically obtain UML diagrams from the modelled specifications, for example, using ATL[4] transformations.

. The *Cametamodel* metamodel specifies the required and available entities for the definition of *Communicative Event* Diagrams, Message *Structures* and *Event Specification Templates*.

Next, the general setup operations and tool usage are presented.

# 4. Tool setup

The following sections introduce the current tool development status, downloading links and the installation and uninstallation operations.

## 4.1. Tool status

The current status of the tool is a prototype under a proof of concept phase. Message Structures are automatically transformed into UML class diagrams. It is expected to upgrade the transformation engine to cover UML activity diagrams too.

---

[2] Physical activities such as "A warehouse worker piles up the boxes" where the client which may be relevant enough to model them, but always at a lower level of abstraction, using stepwise refinement mechanisms

[3] Cametamodel stands for CA (Communication Analysis) metamodel

[4] ATLAS Transformation Language

The automatic code generation of the specified systems is not yet covered and requires external tools, such as the *Integranova Model Execution System[5]*.

Tool authors are interested in use cases, beta testing and user experience of using the GREAT platform, currently available as the *GREAT Process Modeller* tool. Specially, their ambition is to assess the process innovation potential of the methodology, and its underlying supporting tool.

## 4.2. Tool download

The *GREAT Process Modeller* product can be downloaded as a compressed bundle (.zip) from:

Not public available yet. Authors should have already provided it to the reader.

## 4.3. Tool installation

The execution requirements are:

- Microsoft Windows operative system; Windows 7 or superior is suggested
- JRE (Java Runtime Environment); version 7 is suggested

To execute the tool you need to:

- Uncompress the .zip bundle to your desired location; i.e. c:\GREAT_product\
- Launch the product; i.e.: c:\GREAT_product\eclipse\GREAT_PM.exe

All the tool-generated data (diagrams, models, etc.) will be stored under the selected workspace (i.e. c:\GREAT_product\GREATspace\) by default.

## 4.4. Tool uninstallation

You can freely remove the uncompressed bundle files and generated data from the installation phase. Remember to keep a copy of the generated data if you are willing to.

You are also free to uninstall JRE if you had to install it during the installation phase.

## 5. Tool use[6]

This section illustrates a set of steps that conform the CA (Communication Analysis) procedure to specify software systems' requirements from the business modeller and analyst perspective.

---

[5] http://www.integranova.com/integranova-m-e-s/

[6] The tool is an Eclipse Java product. The reader should be familiar with some Eclipse basic topics such as Eclipse perspectives and views (i.e. Project Explorer, Outline, Palette, Properties). For further information the reader is encouraged to contact the authors for more precise tool operation.

When the tool is first run, it may already contain some projects (i.e. the ExpenseReport and Superstationery demo examples). You may keep them, remove or move to your desire destinations.

## 5.1. Project creation

Before you can proceed to specify system requirements throughout the GREAT methodology, you will need to create, or use an existing, Eclipse project. To create a new project:

- Go to Eclipse *Project Explorer* View..
- Right click, and select New -> Project ...
- Use the General -> Project Wizard
- Type a project name
- Click Finish

## 5.2. Cametamodel diagram creation

*Communicative Event Diagrams* can be created following the next operations:

- Right click under the newly, or currently available, Eclipse project
- Choose New -> Example
- Select the *Cametamodel Diagram* wizard
- Click Next
- Change the *File name* if you desire, keeping the .cametamodel_diagram extension
- Click Finish

## 5.3. Organisational roles definition

The first step in the system requirements specification is to define the system actors through organisational roles. To do so you will need:

- The tool Palette (as an Eclipse view)
- Blue-box diagram object: ORGANISATIONAL ROLES

You can create new roles by drag&drop of "Organisational Role" from the palette to the blue-box object (into the half bottom part of the box).

You can remove or rename the existing organisational roles.

## 5.4. Communicative Event Diagram definition

For a correct diagram definition you will almost need to:

- Create a Start and End nodes
- Create almost a Communicative Event (tough not required, you may want to include Event variants inside a Communicative Event)

- Create a Primary actor (related to an Organisational role) for each of the Communicative Events. Tough not required, you may also want to create Receiver actors too
- Connect all Primary actors to its corresponding Communicative Events through the Ingoing object available in the palette (Receivers are connected through the Outgoing object)
- You may also need to define some basic diagram flows through And/Or objects available in the palette
- Connect all the Start/End, And/Or, Communicative Events (or Event variants) through their precedence relationships using the Precedence object available in the palette
- You will also need to define all the Message Structures in the diagram. They are attached to the Ingoing/Outgoing objects. Outgoing Message Structures specification is optional.

### 5.5. Message Structures definition

To define a Message Structure, click into the corresponding icon in the Ingoing (green arrow) or Outgoing (red arrow) connection. A tree-based editor is displayed. You can define the contents of the message right clicking under the Message Structure item in the tree. Doing so over the newly created items you can keep defining lower details of the message.

### 5.6. Event Specification Templates definition

This is currently optional in the tool. You may further specify the communicative events using the Eclipse *Properties* view. It can be displayed right clicking on a Communicative Event.

### 5.7. Cametamodel to UML class diagrams transformation

The transformation from the tool diagrams to UML class diagrams is a straightforward procedure:

1) Open your desired diagram, and click the "CA to UML transformation button" in the toolbar.
2) If the transformation went successful, a message dialog should have informed you about a new .uml file creation (file replacement if it already exist) under your current Eclipse project. If the transformation went wrong, a message dialog should indicate you about the error. This is a common case of under-specification of the diagram. Check the *Communicative Event Diagram* and *Message Structure* definition sections for correctness.
3) Right click on the corresponding .uml file and select Initialize Class Diagram
4) Choose a File name (keep the extension untouched) and click Finish
5) The UML class diagram should have been automatically opened